\documentclass[aps,prev,twocolumn,preprintnumbers,floatfix,nofootinbib]{revtex4-1}
\usepackage{graphicx}
\usepackage{bm}
\usepackage{times}
\usepackage{slashed}
\usepackage{color}
\pdfoutput=1

\begin{document}

\preprint{PI/UAN-2019-662FT,\,\, IIPDM-2019}
\title{Light Dark Matter: A Common Solution to the Lithium and $\boldsymbol{H_0}$ Problems}
\author{Jailson Alcaniz$^{a}$}\email{alcaniz@on.br}
\author{Nicolás Bernal$^{b}$}\email{nicolas.bernal@uan.edu.co}
\author{Antonio Masiero$^{c,d}$}\email{antonio.masiero@pd.infn.it}
\author{Farinaldo S. Queiroz$^{e}$}\email{farinaldo.queiroz@iip.ufrn.br}

\affiliation{$^{a}$~Observat\'orio Nacional, 20921-400, Rio de Janeiro, RJ, Brasil\\
$^b$~Centro de Investigaciones, Universidad Antonio Nari\~no, Bogot\'a, Colombia\\
$^c$~Dipartimento di Fisica e Astronomia ``G. Galilei'', Universita di Padova, Italy\\$^d$~Istituto Nazionale Fisica Nucleare, Sezione di Padova, I-35131 Padova, Italy\\
$^e$~International Institute of Physics,
Universidade Federal do Rio Grande do Norte, Campus Universit\'ario, Lagoa Nova, Natal-RN 59078-970, Brasil
}

\vspace{1cm}

\begin{abstract}
Currently, the standard cosmological model faces some tensions and discrepancies between observations at early and late cosmological time.  One of them concerns the well-known $H_0$-tension problem, i.e., a $\sim4.4\sigma$-difference between the early-time estimate and late-time measurements of the Hubble constant, $H_0$. Another puzzling question rests in the cosmological lithium abundance, where again local measurements differ from the one predicted by Big Bang Nucleosynthesis (BBN). In this work, we show that a mechanism of light dark matter production might hold the answer for these questions. If dark matter particles are sufficiently light and a fraction of them were produced non-thermally in association with photons, this mechanism has precisely what is needed to destroy Lithium without spoiling other BBN predictions. Besides, it produces enough radiation that leads to a larger $H_0$ value, reconciling early and late-time measurements of the Hubble expansion rate without leaving sizable spectral distortions in the Cosmic Microwave Background spectrum. 
\end{abstract}

\maketitle

\paragraph*{{\it Introduction.}}

The standard $\Lambda$-Cold Dark Matter ($\Lambda$CDM) model provides a successful description of the structure and evolution of the universe from its early stages to the present. However, as cosmological observations increase in number and accuracy -- some of the current constraints on the cosmological parameters can reach now sub-percent level -- tensions and discrepancies between early and late time cosmic evolution emerged, requiring either a better understanding of the systematic errors or an extension of the standard model or the discovery of new physics -- we refer the reader to \cite{Verde:2019ivm} for a recent review.

One of these tensions concerns the current difference between the early-time estimates and late-time measurements of the Hubble constant, $H_0$ ($\equiv 100h$ ${\rm{km/s/Mpc}}$). Using distance measurements of galaxies in the local universe calibrated by Cepheid variables and type Ia Supernovae (SNe Ia), Riess {\it{et al.}}~\cite{Riess:2019cxk} reported $H_0 = 74.03 \pm 1.42$ ${\rm{km/s/Mpc}}$, which differs by $\sim 4.4\sigma$ from the latest Cosmic Microwave Background  (CMB) estimate assuming the standard model, $H_0 = 67.36 \pm 0.54$ ${\rm{km/s/Mpc}}$~ \cite{Aghanim:2018eyx}. Such a discrepancy can reach $\sim 5.8\sigma$ when other late-time measurements of $H_0$ \cite{Wong:2019kwg,2019ApJ...882...34F} are combined, reinforcing what is known nowadays as the $H_0$-tension problem  \cite{Verde:2019ivm}. 

Another ongoing issue in the standard cosmology concerns a long-standing problem of the Lithium ($^7Li$) abundance, known as the Lithium problem~\cite{Sbordone:2010zi}. 
It is well known that during the first few minutes of the universe lifetime, nuclear reactions took place and produced light elements such as $^4He$, $D$, $^3He$ and $^7Li$. Using standard calculations one can estimate the abundance of these elements and find good agreement with the observations up to the baryon-to-photon ratio, $\eta$, which is extracted from CMB measurements \cite{Coc:2017pxv,Singh:2019myn}. Moreover, by fixing  $\eta$, no free parameter is left in the Big Bang Nucleosynthesis (BBN) calculations and the primordial abundances of these elements are thus affected only by tamed uncertainties in the nuclear cross-section \cite{Damone:2018mcf}. It is also remarkable that the calculated primordial abundances nicely match the astronomical observations. Albeit, there is a puzzling discrepancy in the $^7Li$ abundance by roughly a factor of three. The predicted abundance reads $^7Li/H = 4.68 (\pm 0.67) \times 10^{-10}$ \cite{Cyburt:2015mya} while measurements point to $^7Li/H = 1.58 (\pm 0.31) \times 10^{-10}$ \cite{Coc:2017pxv}. This difference between the calculated and observed Lithium primordial abundances has been reconciled neither by new observations nor by nuclear physics. Therefore, the long-standing Lithium problem casts doubt either on the success of BBN or on our understanding of the thermal history of the universe.  
  
In the past few years, several attempts have been put forth trying to solve individually one of these problems. Some of them invoked connections to the cosmological dark sector (see e.g. \cite{Vattis:2019efj} and references therein). In many cases, they end up altering too much the abundances of the other light elements or distorting the CMB power spectrum. It has been noticed that an increase in the number of relativistic degrees of freedom is a viable solution to the $H_0$ problem and that late electromagnetic injection may solve the Lithium problem. Aren't these two problems connected? Aren't they connected to dark matter? Instead of trying to solve these two problems individually we explore a common origin. Indeed, in this work we present a possible unified solution to both problems via the production of light dark matter particles in the early universe. It is known that the bulk of the dark matter particles cannot be relativistic at the matter-radiation equality for the sake of structure formation. However, considering that only a small fraction of it is relativistic, we show that both Lithium and $H_0$ problems may be solved. 

\begin{figure}[t!]
\centering
\includegraphics[width=\columnwidth]{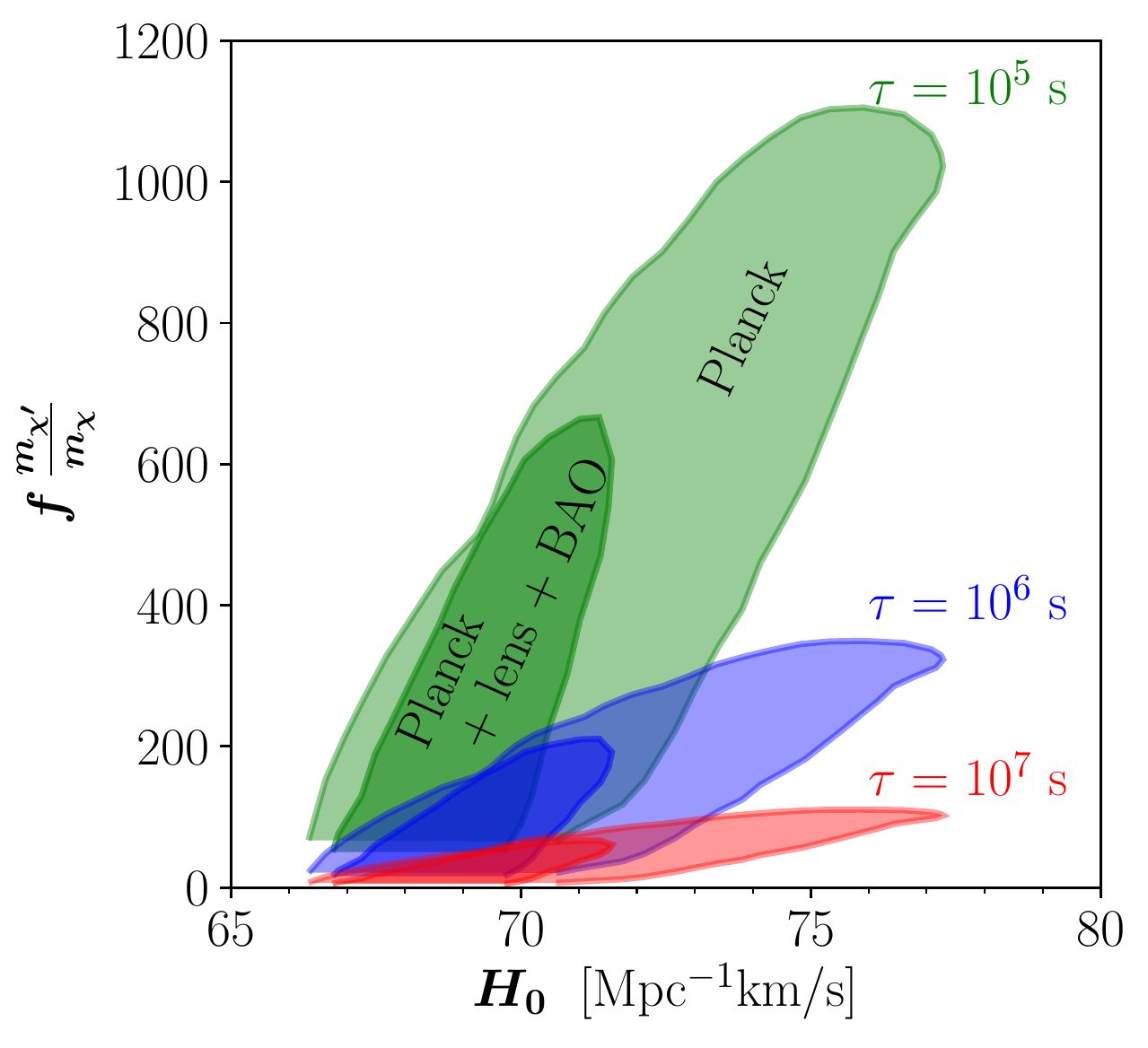}
\caption{Effect of the dark matter production mechanism on $H_0$ for lifetime $\tau=10^5$~s (green), $\tau=10^6$~s (blue) and $\tau=10^7$~s (red).
The contours correspond to 95\% CL constraints from Planck 2015 power spectra excluding high $l$ polarisation (light colors) and adding CMB lensing and BAO (dark colors).}
\label{plot1}
\end{figure}

\paragraph*{{\it  Light Dark Matter and $H_0$.}} In order to understand how the production of light dark matter can lead to larger values of $H_0$ one needs to go over a few steps. If dark matter, $\chi$, is produced via a two body decay, i.e., $\chi^\prime \rightarrow \chi +\gamma$, then energy and momentum conservation implies  $E_{\chi^\prime} =E_\gamma + m_\chi \gamma_\chi$ and $p_{\chi^\prime}=E_\gamma + p_\chi$. Now, assuming that $\chi^\prime$ decays at rest we obtain $\gamma_\chi = m_{\chi^\prime}/2 m_\chi + m_\chi/2m_{\chi^\prime}$. In the radiation-dominated era, $a=(t/t_0)^{1/2}$, thus the prompt decay of $\chi^\prime$ yields
\begin{equation}
\gamma_\chi(t) = \sqrt{\frac{\left(m_{\chi'}^2-m_\chi^2\right)^2}{4\,m_{\chi'}^2\,m_\chi^2}\,\frac{\tau}{t}+1}\,,
\end{equation}
which represents the boost factor of the dark matter particles as a function of time in a radiation-dominated universe, and where $\tau$ corresponds to the time at which $\chi'$ instantaneously decays. We emphasize that this equation is strictly valid only in the ultra-relativistic case, which happens for $m_{\chi^\prime} \gg m_\chi$. Our goal here is to illustrate the framework. 
The dark matter number density scales with $a^{-3}$, which implies  $\rho_\chi(a) = \rho_c\,\Omega_\chi \,a^{-3}$, whereas relativistic neutrinos density evolves as $\rho_\nu(a) = \rho_c\, \Omega_\nu\, a^{-4}\,N_\nu/3$, where $N_\nu$ is the number of neutrino species. Hence, at the matter-radiation equality, $a_{\rm eq} \sim 3 \times 10^{-4}$, one neutrino species represents about 16\% of the dark matter density \cite{Hooper:2011aj}. In other words, if dark matter particles had a boost of $\sim 16$\% they could mimic the effect of neutrino species, increasing the Hubble rate.  In this way, the additional radiation density is given by $\rho_{\rm extra}=f \times \rho_{\chi} (\gamma_{\chi} -1)$, where $f$ accounts for the fraction of the dark matter particles produced via this mechanism, with 
 $\Delta N_{\rm eff}\equiv N_{\rm eff} -N_{\rm eff}^{\rm SM}= \rho_{\rm extra}/\rho_{1\nu}$, $N_{\rm eff}^{\rm SM}$ being the number of relativistic degrees of freedom in the $\Lambda$CDM model. Using the relations above we obtain, 
\begin{equation}
\Delta N_{\rm eff}  = f\times (\gamma_\chi -1)/0.16,
\end{equation}
which implies
\begin{eqnarray}
\Delta N_{\rm eff} &  \sim & 5 \times 10^{-3}\left( \frac{\tau}{10^6~{\rm s}} \right)^{1/2}\nonumber\\ 
                     &   &
  \times \left[\left( \frac{ m_{\chi^{\prime}} }{2 m_{\chi}} + \frac{m_{\chi}}{2 m_{\chi^{\prime}}} -1 \right) \right]\times f,
\label{eqNeff1}
\end{eqnarray}
for $t=t_{\rm CMB}$.
Therefore, if a fraction of the dark matter particles are non-thermally produced, with $m_{\chi^\prime} \gg m_\chi$, they generate extra relativistic degrees of freedom. Several works have assessed the impact of $N_{\rm eff} >3$ on the CMB power spectrum, as $N_{\rm eff}$ and $H_0$ are positively correlated \cite{Ade:2013zuv,Karwal:2016vyq,Bernal:2016gxb,Graef:2018fzu}. 
We translate this correlation to our framework using Eq.~\ref{eqNeff1} and find the results displayed in Fig.~\ref{plot1}. Clearly, for $150 < f\, m_\chi^\prime/m_\chi  < 350$ and $\tau=10^6$~s, we may obtain $H_0 \sim 74$~Mpc$^{-1}$km/s. As mentioned earlier, the fraction of dark matter produced via this mechanism is expected to be small since the bulk of the dark matter should be non-relativistic at the matter-radiation equality. 

To illustrate how that works we need to invoke the concept of free-streaming, which is the scale below which a dark matter particle do not cluster. Therefore on sufficiently
small scales, the growth of perturbations is suppressed if the dark matter production mechanism discussed here were the main one. We can estimate the free streaming of the dark matter particles, produced via this mechanism with $m_{\chi^\prime} \gg m_\chi$ to be \cite{Cembranos:2005us}, 
\begin{equation}
\lambda_{\rm FS} \sim 1~{\rm Mpc}\left( \frac{\tau}{10^6~{\rm s}}\right)^{1/2}\left(\frac{m_{\chi^\prime}}{2 m_\chi} - \frac{m_\chi}{2 m_{\chi^\prime}}\right),
\end{equation}
which leads to a $\lambda_{\rm FS}$ much larger than a Mpc scale, in complete disagreement with the Lyman-alpha forest that sets $\lambda_{\rm FS} \ll 1$~Mpc \cite{Viel:2010bn}. We therefore conclude that such a production mechanism cannot be responsible for the overall dark matter density. In order to have a more quantitative assessment about the fraction $f$ of dark matter particles that can be produced in this way we need to study the matter power spectrum \cite{Smith:1997wq,Eisenstein:1997jh}. 

In the matter-dominated phase of the universe, the linear fluctuation evolves below the free-steaming length as, $\delta \propto a^{1-3/5f}$ \cite{Ma:1996za} for $f\ll 1$. Hence, the suppression that such production mechanism above brings to the matter-power spectrum is quantified by $\Delta =\frac{\delta_{f\neq0}}{\delta_{f=0}}=(a_{\rm eq}/a)^{-3f/5}\sim \exp(-4.9 f)$. Considering current measurements of the amplitude of matter clustering, $\sigma_8$ \cite{Abbott:2017wau,Zennaro:2017qnp,Parimbelli:2018yzv,Giusarma:2018jei}, CMB data \cite{Aghanim:2018eyx}, the  clustering power in the Lyman-alpha forest \cite{Viel:2010bn}, and the results of cosmological simulations \cite{Liu:2017now,Fidler:2018bkg,Zennaro:2019aoi} one needs a nearly pure cold dark matter regime. In other words, we require $\Delta > 0.95$, which implies that $f< 0.01$. Therefore, only a small fraction of dark matter may be produced via the proposed mechanism, in order to be consistent with large-scale structure observations. Interestingly, although small it provides a possible solution to the $H_0$-tension problem discussed above. Generally speaking $N_{\rm eff}$ could be generated via new light thermal relics and new interactions with the active neutrinos etc \cite{Anchordoqui:2011nh}. In our work, the increase in $N_{\rm eff}$ and consequently in $H_0$ are tied to the dark matter production mechanism.  and many things, such new light species coupled In what follows, we discuss the effect of this dark matter production on the long-standing $^7Li$ problem. 

\paragraph*{{\it Light Dark Matter and the Lithium Problem.}}

The primordial Lithium abundance is derived from low-metallicity halo stars observations \cite{Sbordone:2010zi}, whereas the theoretical prediction is based on the baryon-to-photon ratio, which is extracted from CMB observations. These determinations of the $^7Li$ abundance differ by roughly a factor of three~\cite{Sbordone:2010zi}, and the fact that nuclear reactions, cosmological modifications, modified statistics, possible primordial magnetic fields, and exotic decays involving charged particles all failed to solve the Lithium problem represents an incomplete success for cosmology, as reviewed in \cite{Cyburt:2015mya}. Some solutions to this problem have been proposed in the past years (see e.g. \cite{Goudelis:2015wpa} and references therein). 

In a three-body decay the photon in the final state would have a continuous spectrum with energy sufficiently large to alter the abundance of several elements since these photons would induce numerous nuclear reactions. On the other hand, having an arbitrary two-body decay with high energetic photons does not suffice either because secondary nuclear reactions occur. For instance, taking our scenario as an example, in the two-body decay with $m_{{\chi}^\prime} \gg m_{\chi}$, photons are emitted with $E_\gamma\simeq m_{\chi^\prime}/2$, electromagnetic cascades would be induced by the injection of an energetic $\gamma$ in a medium comprised of radiation, magnetic fields and matter \cite{Iocco:2008va}.  Such photon emission could be catastrophic to the abundance of light elements because these photons may induce nuclear reactions and alter the BBN predictions, which is precisely the problem with past proposals. 

\begin{figure}[!t]
\centering
\includegraphics[width=\columnwidth]{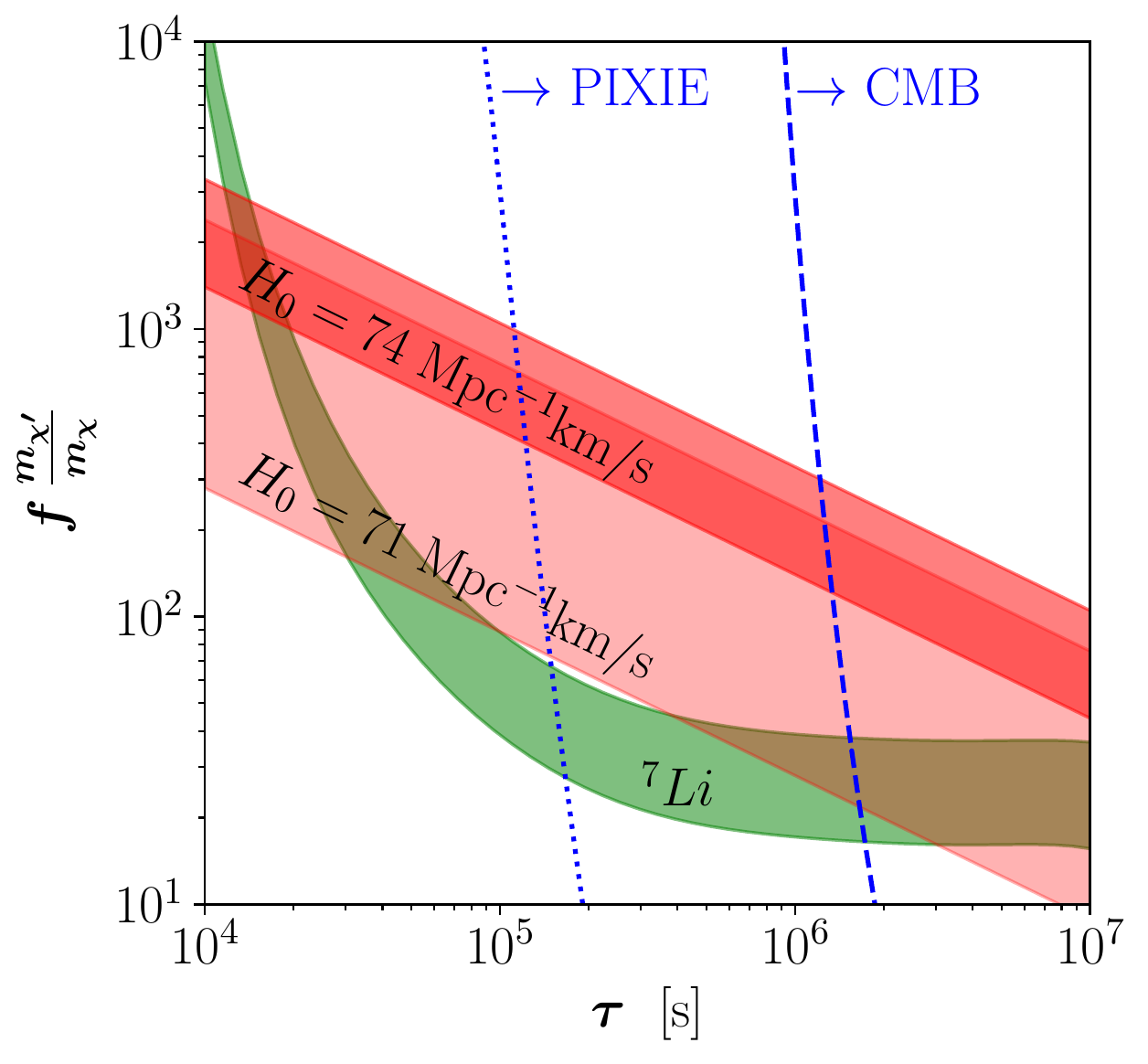}
\caption{Region of parameter space in which the Lithium abundance is diluted by 20-80\% from BBN original prediction (green), and where $H_0=71$~Mpc$^{-1}$km/s (light red) or $H_0=74$~Mpc$^{-1}$km/s (dark red).
The blue curves correspond to current (dashed) and projected (dotted) CMB spectral distortion bounds.
For $\tau\simeq2\times 10^4$~s one can simultaneously accommodate the Lithium and $H_0$ problems with light dark matter.
Notice if one assumes $f\ll 1$, sub-keV dark matter arises.
See text for details.}
\label{plot2}
\end{figure}

However, if the energy of the photon injected is small enough, between $1.59-2.22$~MeV the electromagnetic cascades develop differently, allowing to destroy enough $^7Li$ without affecting the abundance of other elements. This energy window has to do with the Beryllium nucleus that has a photon dissociation threshold of $1.59$~MeV and Deuterium whose threshold is $2.22$~MeV. Taking into account the fact that MeV photons induce non-thermal nucleosynthesis processes where the final spectrum of the photons is significantly different from the injected one it has been shown that one could successfully solve the Lithium problem via electromagnetic injection \cite{Poulin:2015woa}. Now, knowing that the total electromagnetic energy released is $\zeta_{\rm em}=E_\gamma\,Y_\gamma$, where $Y_\gamma$ is the ratio of the number density of photons produced in the decay to the number density of CMB photons, which yields,
\begin{equation}
\zeta_{\rm em} \sim  1.5\times10^{-8}\ \mbox{MeV} \left( \frac{m_{\chi^{\prime}}}{m_{\chi}} -\frac{m_\chi}{m_{\chi^\prime}} \right)\left( \frac{f}{0.01}\right),
\label{emEq}
\end{equation}
we can obtain the region in the parameter space, lifetime vs $\zeta_{\rm em}$, in which the Lithium abundance is reduced from 20\% up to 80\% for instance. Using Eq.~\ref{emEq} we may  display this region in terms of $f m_{\chi^\prime}/m_\chi$ vs lifetime.
Considering the region of the parameter space favored in Fig.~\ref{plot1}, which fix $f m_{\chi'}/m_\chi$, we assess whether is possible to reconcile both the $H_0$ and the Lithium problems, as $f m_{\chi'}/m_\chi$ governs the overall energy injection as well. The results are shown in Fig.~\ref{plot2} which shows in green the region of parameter space in which the $^7Li$ abundance is diluted in 20\%-80\%. As mentioned earlier, the photon should be emitted with an energy of about 2~MeV. As our reasoning is built under the assumption that $m_{\chi}\gg m_\chi$, then $m_{\chi^\prime}=4$~MeV throughout. One can easily fix the fraction of dark matter produced in this mechanism, $f$, to extract the dark matter mass. For instance, with $f m_{\chi^\prime}/m_{\chi} = 10^3$, we get $m_{\chi}= 0.04$~keV for $f=0.01$.

In Fig.~\ref{plot2} the dashed (dotted) blue curves correspond to the limits based on the CMB spectral distortion. An additional energy injection in the form of photons brings distortions to the CMB black-body spectrum. For $\tau < 10^3$~s, these photons produced in the decay quickly thermalize via processes such as Compton scattering $e\gamma \rightarrow e\gamma$ and double Compton scattering $e\gamma \rightarrow e\gamma\gamma$ \cite{Feng:2003xh}. Although, for  larger lifetimes these processes are not sufficiently efficient and end up altering the photon spectrum of the CMB to a Bose-Einstein distribution with a chemical potential $(\mu \neq 0)$. Therefore, constraints on a chemical potential, $\mu < 9\times 10^{-5}$, lead to bounds on the total energy injection \cite{Chluba:2011hw} which is related to $ m_{\chi^\prime}/m_\chi$, as displayed in Fig.~\ref{plot2}.
The dotted blue curve represents a projection using the PIXIE setup corresponding to $\mu < 5 \times 10^{-8}$ \cite{Kogut:2011xw}. These CMB spectral distortion bounds exclude the region at the right of the curves.  



\paragraph*{{\it Discussion.}}

Fig.~\ref{plot2} also shows that there is a region of the parameter space corresponding to $\tau\simeq 2\times 10^4$~s and $f\,m_{\chi'}/m_\chi\simeq 2\times 10^3$ which is capable of diluting the Lithium abundance while simultaneously reproducing $H_0\simeq 74$~Mpc$^{-1}$km/s, as indicated by recent local measurements --  while being consistent structure formation and CMB constraints.
These results rely on the existence of very light dark matter with sub-keV mass, with a small fraction of them being produced in association with photons.
In other words, the production of sub-keV dark matter seems to provide a possible solution to current problems in the standard cosmology. We emphasize that our results do depend on the dark matter abundance. Thus, in order to solve both problems dark matter must appear in the final state of the decay channel proposed. It could not have been any particle, as the abundance of dark matter determines the amount of extra radiation in the early universe which is directly associated with the increase in $H_0$. Therefore, if new independent measurements of $H_0$ converge increasing the need for new physics,  sub-keV dark matter stands as a viable possibility, strengthening the need for new detectors sensitive to very light dark matter \cite{Battaglieri:2017aum}. 


Finally, sub-keV dark matter could have been produced non-thermally in the early universe via the freeze-in mechanism~\cite{Bernal:2017kxu}.
In fact, if the production channel has a small available phase-space, the dark matter free-streaming length is suppressed and therefore the Lyman-alpha constraint can be avoided~\cite{Heeck:2017xbu, Boulebnane:2017fxw}.
A very light dark matter could also be produced via the misalignment mechanism typical of axion dark matter models~\cite{Marsh:2015xka}. It is also important to mention that the Fermi-Dirac statistics limits dark matter mass to be higher of few hundreds of eV, independent of the production mechanism~\cite{Tremaine:1979we, Boyarsky:2008ju}. Therefore, a sub-keV dark matter model is perfectly conceivable once we invoke non-thermal processes. Moreover, as a small fraction of dark matter may be produced via this mechanism larger-scale structures are not affected \cite{Bringmann:2018jpr}.

\paragraph*{{\it Final Remarks.}} Recently, several proposals have surfaced trying to induce a larger expansion rate  via an increase on $N_{\rm eff}$, such as dark matter interactions with active neutrinos or via the introduction of new light species. Moreover, numerous attempts to solve the Lithium have been put forth in the context of late particle decays involving particles with an electric charge, neutral particles and even decays having dark matter-like particles in the final state. Often times dark matter in these works refer to neutral particles which are stable and interact weakly. However,  being stable and a weakly interacting massive particle does not warrant a particle the role of dark matter in our universe. Our findings are tied to the dark matter density instead. 

In this work, we go beyond that and propose a possible, unified solution to  the Lithium and $H_0$-tension problems. Such result  is obtained through a mechanism of light (sub-keV) dark matter production in which a fraction of the dark matter particles are produced non-thermally in association with photons. We have discussed the observational viability of this framework in light of the current large-scale structure and CMB data and shown that for an interval of the parameter values the Lithium abundance may be significantly reduced without altering other elements abundance, while simultaneously increasing the current expansion rate in agreement with current local measurements.

As is well know, light dark matter has gained attention recently as the still popular WIMPs (Weakly Interacting Massive Particles) subject to intense experimental searches have not been observed. We believe that the results discussed in this paper further motivate experimental searches for sub-keV dark matter.

\section*{Acknowledgments}
The authors are grateful to Francesco D'Eramo, Marco Peloso, Clarissa Siqueira and Pasquale Serpico for comments. JSA acknowledges support from CNPq (grant Nos. 310790/2014-0 and 400471/2014-0) and FAPERJ (grant No. E-26/203.024/2017). FSQ thanks CNPq grants 303817/2018-6 and 421952/2018-0, and ICTP-SAIFR FAPESP grant 2016/01343-7 for the financial support.
NB is partially supported by Spanish MINECO under Grant FPA2017-84543-P. This project has received funding from the European Union's Horizon 2020 research and innovation program under the Marie Sklodowska-Curie grant agreements 674896 and 690575, and from Universidad Antonio Nariño grants 2018204, 2019101 and 2019248.  The research of AM is supported by the ERC Advanced Grant No. 267985 (DaMeSyFla) and by the INFN.
We thank the High Performance Computing Center (NPAD) at UFRN for providing computational resources.

\bibliographystyle{reffixed}
\bibliography{darkmatter}

\end{document}